\documentclass[11pt]{book}

\usepackage{amsmath}
\usepackage{amsfonts}
\usepackage{epsfig}

\newcommand{\cc}[1]{{#1}^{\!\!*}} 

\newcommand{\C}{ \mathbb{C} }

\newcommand{\rme}{ \mathrm{e} }
\newcommand{\rmi}{ \mathrm{i} }
\newcommand{\rmd}{ \mathrm{d} }
\newcommand{\rmD}{ \mathrm{D} }

\newcommand{\tr}{\mathop{\mathrm{Tr}}}
\newcommand{\Pfaff}{\mathop{\mathrm{Pfaff}}}
\newcommand{\sgn}{\mathop{\mathrm{sgn}}}
\newcommand{\erfc}{\mathop{\mathrm{erfc}}}
\newcommand{\diag}{\mathop{\mathrm{diag}}}
\newcommand{\tridiag}{\mathop{\mathrm{tridiag}}}
\newcommand{\re}{\mathop{\mathrm{Re}}}
\newcommand{\im}{\mathop{\mathrm{Im}}}

\makeatletter

\renewenvironment{thebibliography}[1]
     {\section*{\bibname}%
      \@mkboth{\MakeUppercase\bibname}{\MakeUppercase\bibname}%
      \list{\@biblabel{\@arabic\c@enumiv}}%
           {\settowidth\labelwidth{\@biblabel{#1}}%
            \leftmargin\labelwidth
            \advance\leftmargin\labelsep
            \@openbib@code
            \usecounter{enumiv}%
            \let\p@enumiv\@empty
            \renewcommand\theenumiv{\@arabic\c@enumiv}}%
      \sloppy
      \clubpenalty4000
      \@clubpenalty \clubpenalty
      \widowpenalty4000%
      \sfcode`\.\@m}
     {\def\@noitemerr
       {\@latex@warning{Empty `thebibliography' environment}}%
      \endlist}

\makeatother

\newcommand{\sect}[1]{\setcounter{equation}{0}\section{#1}}

\renewcommand\bibname{References}

\begin{document}

\setlength{\baselineskip}{5.0mm}


\setcounter{chapter}{17}
\chapter[Non-Hermitian Ensembles]{Non-Hermitian Ensembles}
\thispagestyle{empty}

\ \\

\noindent
{{\sc Boris A Khoruzhenko}$^1$ and {\sc Hans-J\"{u}rgen Sommers}$^2$
\\~\\$^1$ Queen Mary University of London, School of Mathematical Sciences,\newline London E1 4NS, UK
\\
$^2$Fachbereich Physik, Universit\"{a}t
Duisburg-Essen, 47048 Duisburg, Germany }

\begin{center}
{\bf Abstract}
\end{center}
This is a concise review of the complex, real and quaternion real Ginibre random matrix ensembles and their elliptic deformations. Eigenvalue correlations are exactly reduced to two-point kernels and discussed in the strongly and weakly non-Hermitian limits of large matrix size.

\sect{Introduction}\label{intro}

The study of eigenvalue statistics in the complex plane was initiated in 1965 by Ginibre \cite{Gin65} more than 40 years ago who introduced a three fold family of Gaussian random matrices (complex, real and quaternion real) as a mathematical extension of Hermitian random matrix theory. Although no physical applications of the theory were in sight at that time, Ginibre expressed the hope that `the methods and results will provide further insight in the cases of physical interest or suggest as yet lacking applications'. Nowadays, statistics of complex eigenvalues have many interesting applications in modeling of a wide range of physical phenomena. They appeared in the studies of quantum chromodynamics (see Chap. 32), dissipative quantum maps \cite{Gro88} and scattering in chaotic quantum systems (see Chap. 34), growth processes (see Chap. 38), fractional quantum-Hall effect \cite{DiF94} and Coulomb plasma \cite{For97}, stability of complex biological \cite{May72} and neural networks \cite{Som88}, directed quantum chaos in randomly pinned superconducting vortices \cite{Efe97}, delayed time series in financial markets \cite{Kwa06},  random operations in quantum information theory \cite{Bru09}, and others.

This Chapter gives an overview of the three Ginibre ensembles and their elliptic deformations. We tried to keep our exposition self-contained providing hints of derivations. Some of the derivations included are new. We did not have space to cover the chiral extensions of the Ginibre ensembles, only mentioning briefly the chiral companion of the complex Ginibre ensemble \cite{Osb04}. This topic is partly covered in Chap. 32. The real and quaternion real (qu-r) companions were solved only recently, see \cite{Ake09b} and \cite{Ake05}. There are also important non-Hermitian ensembles of random matrices relevant in the context of quantum chaotic scattering which are only mentioned here (but see Chap. 34 for a summary of results and the survey paper \cite{Fyo03} for details). Also, we will not discuss the complete classification of non-Hermitian matrix ensembles depending on the action of a few number of involutions \cite{Ber02,Mag08}.

On a macroscopic scale, all three Ginibre ensembles exhibit similar patterns of behavior with a uniform distribution of eigenvalues and sharp (Gaussian) fall in the eigenvalue density when one transverses the boundary of the eigenvalue support. The similarities extend to the microscopic scale as well but only away from the real line where all three ensembles exhibit a cubic repulsion of eigenvalues. In the vicinity of real line and on the real line their behavior is very different due to the differences in symmetries. The eigenvalue correlation functions have either determinantal (complex) or pfaffian (real and qu-r) form with the kernel being almost identical far away from the real line, again the differences coming from a pre-exponential factor describing the transition from the real line into the bulk of the spectrum.

One expects the eigenvalue statistics provided by the Ginibre ensembles to be universal within their symmetry classes. Establishing such universality is an open and challenging problem, especially for the real and qu-r ensembles. There is some evidence for universality in the complex case where every solved model led to Ginibre's form of correlations, including complex normal matrices where the universality of Ginibre's correlations was proved in a general class of matrix distributions \cite{Ame08}. This result can be applied to complex matrices (\ref{eq-b:1a}) as for this class of ensembles the induced eigenvalue distribution does not differ from the one for normal matrices \cite{Oas97}, see also Chap. 38.

\sect{Complex Ginibre Ensemble} \label{sec2}
{\bf Measure, change of variables} \quad The complex Ginibre ensemble is defined on the space of complex $N\times N$ matrices by the probability measure
\begin{equation}\label{eq-b:1}
\rmd\mu (J)= \exp\big(-\tr JJ^{\dagger} \big) |\rmD J|\, .
\end{equation}
Here $\rmD J=\prod_{i,j=1}^N (\rmd J_{ij}\rmd \cc{J_{ij}}/2\pi)$ is the (exterior) product of the one-forms in matrix entries and $|\rmD J|$ is the corresponding Cartesian volume element. With probability one, the matrix $J$ has $N$ distinct eigenvalues $z_j$. On ordering the eigenvalues in an arbitrary but fixed way, one can think of them as of random variables. The corresponding joint probability distribution function (jpdf) can be obtained by changing variables in (\ref{eq-b:1}). This can be done in several ways, we outline here a calculation due to Dyson \cite{Meh04}.

On making use of the Schur decomposition, $J$ can be brought to triangular form $J=U(\Lambda+\Delta)U^{-1}$. Here $U$ is unitary, $\Lambda=\diag (z_1, \ldots, z_N)$ and $\Delta$ is strictly upper-triangular. It is apparent that $U$ can be restricted to the space of right cosets ${\rm U}(N)/{\rm U}(1)^{N}$.
The variations in $J$ are related to those in $\Lambda, \Delta$ and $U$ by the formula $U^{-1}\rmd JU=\rmd \Lambda +\rmd \Delta + \rmd M$, where
\[
(\rmd M)_{ij}=(U^{-1}\rmd U)_{ij}(z_j-z_i)+\sum_{k<j}(U^{-1}\rmd U)_{ik}\Delta_{kj}-\sum_{l>i}\Delta_{il}(U^{-1}\rmd U)_{lj}\, .
\]
The volume form $\rmD J$ does not change on conjugation by unitary matrices. Hence $\rmD J=\rmD (\Lambda+\Delta+M)$ and one gets the Jacobian of the coordinate transformation by multiplying through the matrix entries of $\rmd\Lambda +\rmd\Delta + \rmd M$ using the calculus of alternating differential forms. This leads to the important relation
\begin{equation}\label{eq-b:2}
\rmd\mu (J) = C_N\, \rme^{-\tr (\Delta \Delta^{\dagger} + \Lambda\Lambda^{\dagger} )} \prod_{1\le i<j\ \le N} |z_i-z_j|^2\, |\rmD U|\, |\rmD \Lambda|\, |\rmD\Delta|\, ,
\end{equation}
where $\rmD U=\prod_{i<j} (U^{-1}\rmd U)_{ij}(U^{-1}\rmd U)_{ji}$ is a volume form on the coset space. The density function on the rhs is symmetric in the eigenvalues of $J$. It does not depend on $U$ and is Gaussian in $\Delta$, and these two variables can be easily integrated out.  Thus, if $f(z_1, \ldots, z_N)$ is symmetric in eigenvalues of $J$ then \cite{Gin65}
$\int f \rmd\mu = \int \rmd^2z_1 \cdots \int \rmd^2 z_N P_N(z_1, \ldots, z_N) f(z_1, \ldots, z_N)$
where $\rmd^2z=|\rmd z\rmd z^*|/2$ is the element of area in the complex plane and
\begin{equation}\label{eq-b:4}
P(z_1, \ldots, z_N)=\frac{1}{\pi^N \prod_{j=0}^N j!}\ \rme^{-\sum_{j=1}^N |z_j|^2} \prod_{1\le i<j\ \le N} |z_i-z_j|^2
\end{equation}
is the \emph{symmetrized} jpdf of the eigenvalues of $J$.

In contrast to the Hermitian matrices, the above calculation of the jpdf is not easily extended to other invariant distributions. E.g., it breaks down if $JJ^{\dagger}$ in (\ref{eq-b:1}) is replaced by a higher order polynomial in $JJ^{\dagger}$ \cite{Fei97}
as $U$, $\Delta$ and $\Lambda$ do not decouple in that case, although it still works well for (see Chap. 38)
\begin{equation}\label{eq-b:1a}
\rmd\mu (J) \propto \exp\big(- \tr JJ^{\dagger} - \re \tr \Phi (J)\big) |\rmD J|\, ,
\end{equation}
with $\Phi (J)$ being a potential that ensures existence of the normalisation integral. We will discuss in detail the special case of (\ref{eq-b:1a}) with $\Phi (J)=J^2$ later.

{\bf Correlation functions and orthogonal polynomials} \quad The eigenvalue correlation functions are marginals of the symmetrized jpdf,
\begin{equation}\label{eq-b:4a}
R_n(z_1, \ldots, z_n) = \frac{N!}{(N-n)!} \int {\rm d}^2z_{n+1} \ldots \int {\rm d}^2 z_N P(z_1, \ldots, z_N)\, ,
\end{equation}
normalized to $\int {\rm d}^2z_1\ldots \int {\rm d}^2 z_n\ R_n(z_1, \ldots,  z_n) =N(N-1)...(N-n+1)$. The one-point correlation function is just the density of eigenvalues $\rho (z) =\sum_j \delta^{(2)} (z-z_j)$ averaged over the ensemble distribution, $R_1(z) =\langle \rho(z) \rangle$, so that if $n_D$ is the number of eigenvalues in $D$ then
$\langle n_D \rangle = \int_D \rmd^2 z R_1(z)$.

The eigenvalue correlation functions for the complex Ginibre ensemble can be found in a closed form. The corresponding calculation is almost identical to that for Hermitian ensembles. For the purpose of future reference we shall outline it in a slightly more general setting.

Let $p_m(z)$ be the monic orthogonal polynomials associated with weight function $w(z)\ge 0$ in the complex plane,  i.e.,
$\int \rmd^2 z\ w(z)\, p_m(z)p_n(z)^*= h_n \delta_{m,n}$ and $p_m(z)=z^m+\ldots $.
Then
\begin{equation}\label{eq-b:5}
P(z_1, \ldots, z_N)=\frac{1}{N!\prod_{l=0}^{N-1}h_l} \prod_{j=1}^N w(z_j) \prod_{i<j} |z_i-z_j|^2
\end{equation}
is a probability density in $\C^N$ symmetric with respect to permutations of $z_j$.
Recalling the Vandermonde determinant
$\prod_{1\le i<j\le N} (z_i-z_j)=\det (z^{N-j}_i)_{1\le i,j\le N}=\det (p_{N-j}(z_i))_{1\le i,j\le N}$,
one obtains the important determinantal representation of the jpdf:
\begin{equation}\label{eq-b:6}
P(z_1, \ldots, z_N)=\frac{1}{N!}\det (K(z_i,z_j))_{i,j=1}^N
\end{equation}
with
\begin{equation}\label{eq-b:7}
K(z_1,z_2)= \sqrt{w(z_1)}\sqrt{w(z_2)}\ \sum_{n=0}^{N-1} \frac{p_n(z_1)p_n(z_2)^*}{h_n}\, .
\end{equation}
The kernel $K$ is Hermitian, $K(z_1,z_2)=K(z_2,z_1)^*$, and
$\int \rmd^2  z\ K(z,z)=N$, $\int \rmd^2  z\ K(z_1,z)K(z,z_2)=K(z_1,z_2)$.
Hence by Mehta's `integrating out' lemma \cite{Meh04} (see also the relevant section in Chap. 4)
\begin{equation}\label{eq-b:8a}
R_n(z_1, \ldots, z_n)=\det (K(z_i,z_j))_{i,j=1}^n.
\end{equation}
In particular, the one- and two-point correlation functions are given by  $R_1(z)=K(z,z)$ and $R_2(z_1,z_2)=K(z_1,z_1)K(z_2,z_2)-|K(z_1,z_2)|^2$.

In the complex Ginibre ensemble $w(z)=\rme^{-|z|^2}$. In view of the rotational symmetry the power functions are orthogonal. Thus  $p_n(z)=z^n$,  $h_n=\pi n!$ and
\begin{equation}\label{eq-b:9}
K(z_1,z_2)=\frac{1}{\pi}\, \rme^{-\frac{1}{2}|z_1|^2-\frac{1}{2} |z_2|^2}\sum_{n=0}^{N-1} \frac{(z_1z_2^*)^{n}}{n!}.
\end{equation}
The sum on the rhs is the truncated exponential series and can be expressed in terms of the incomplete Gamma function: $\sum_{l=0}^{N-1} \frac{x^{l}}{l!}= e^{x}\, \Gamma (N, x)/\Gamma (N)$.
The saddle-point integral $\Gamma (N, x)=\int_x^{\infty} \rmd t\  \rme^{-t} t^{N-1}$ comes in handy for asymptotic analysis. We have
\begin{equation}\label{eq-b:11bulk}
R_1(z)=\frac{1}{\pi} \frac{\Gamma(N,|z|^2)}{\Gamma(N)}\ \simeq \ \frac{1}{\pi}\, \Theta (\sqrt{N}-|z|), \quad N\to\infty\, ,
\end{equation}
where $\Theta$ is the Heaviside function, meaning that on average most of the eigenvalues are distributed within the disk $|z|<\sqrt{N}$ with constant density in agreement with the Circular Law \cite{Gir85,Tao08}. The expected number of eigenvalues outside this disk $\simeq \sqrt{\frac{N}{2\pi}}$. The eigenvalue density in the transitional region around the circular boundary, as found from the integral $\Gamma (N,x)$, is given in terms of the complementary error function
\begin{equation}\label{eq-b:11edge}
R_1\big((\sqrt{N}+x\big)\, \rme^{i\varphi})\simeq \frac{1}{2\pi}\erfc (\sqrt{2}\, x), \quad \text{as $ N\to\infty$},
\end{equation}
\cite{For99,Kan05a}. Recalling that
$\erfc (x)\, \simeq\, 2\, \Theta(-x) + \rme^{-x^2}/(\sqrt{\pi}x)$ for large real $|x|$,
one concludes that the density vanishes at a Gaussian rate at the edge.

The kernel in (\ref{eq-b:9}) has a well defined limit as $N\to\infty$ and $|z_{1,2}| =O(1)$,
leading to a simple expression for the correlation functions in this limit \cite{Gin65}
\begin{equation}\label{eq-b:11a}
R_n(z_1, \ldots, z_n)\simeq \frac{1}{\pi^n}\ \rme^{-\sum_j |z_j|^2} \det (\rme^{z_iz_j^*})_{i,j=1}^n\, .
\end{equation}
Equation (\ref{eq-b:11a}) describes eigenvalue correlations at the origin $z_0=0$  and on the (local) scale when the mean separation between eigenvalues is of the order of unity. It also holds true (locally) at any other reference point inside the disk $|z|<\sqrt{N}$, see, e.g. \cite{Bor08}. In particular,
$R_2(z_1, z_2) \simeq \frac{1}{\pi^2}(1-\rme^{-|z_1-z_2|^2})$.
Note that the dependence of $R_n$ on the reference point disappears in the limit $N\to\infty$ and complete homogeneity arises. The eigenvalue correlation functions at the edge of the eigenvalue support can also be found, see \cite{For99,Bor08}.

{\bf Gap probability and nearest neighbor distance} \quad  Consider the disk $D$ of radius $s$ centered at $z_0$ and denote by $\chi_D$ its characteristic function. Define $H(s;z_0)$ to be the conditional probability that given one eigenvalue lies at $z_0$ all others are outside $D$. With $\chi_D$ being the characteristic function of $D$,
\begin{equation}\label{eq-b:12h}
H(s;z_0) = \frac{N}{R_1(z_0)} \int \rmd^2 z_2\cdots \int \rmd^2  z_N  P(z_0,z_2,\ldots, z_N) \prod_{k=2}^N (1-\chi_D(z_k))\, .
\end{equation}
When $s$ gets infinitesimal increment $\delta s$, the decrement in $H$,  $H(s,z_0)-H(s+\delta s,z_0)$, is the probability for the distance between the eigenvalue at $z_0$ and its nearest neighbor to lie in the interval $(s,s+\delta s)$. Therefore $p(s;z_0)=-\frac{\rmd}{\rmd s} H(s;z_0)$ is the density of nearest neighbor distances at $z_0$.

For the complex Ginibre ensemble $H(s;0)$ can be easily computed with the help of the Vandermonde determinant and Andr\'eief-de Bruijn integration formula (Eq. (8.3.41) in Chap. 8). The rotational invariance of the weight function $\rme^{-|z|^2}$ ensures that the monomials $z^m$ stay orthogonal when integrated over the disk $D$, leading to \cite{Gro88}
\begin{equation}\label{eq-b:13}
H(s;0) = \prod_{n=1}^{N-1} \frac{\Gamma(n+1,s^2)}{\Gamma(n+1)}=\prod_{n=1}^{N-1} \left(1-\rme^{-s^2}\sum_{l=n+1}^{\infty}\frac{s^{2l}}{l!}\right)\, .
\end{equation}
The product on the rhs converges quite rapidly in the limit $N\to\infty$ and $H(s;z_0)$ has a well defined limit. In this limit $H(s,0)=1-s^4/2+s^6/6-s^8/24 +O(s^{10})$ for small $s$ and \cite{Gro88} $H(s,0)=\exp [-s^4/4-s^2(\ln s +O(1))]$ for large $s$. Taking the derivative, one gets the cubic law of the eigenvalue repulsion: $p(s;0)= 2s^3+O(s^5)$ for small $s$.

The small-$s$ behavior of $H(s,z_0)$ can also be obtained by expanding the product in (\ref{eq-b:12h}). E.g., on retaining the first two terms  $H(s,z_0)=1-\frac{1}{R_1(z_0)}\int_{|z-z_0|\le s} \rmd^2  z\  R_2(z_0,z) + O(s^4)$. This approach is general and does not rely on the rotational invariance. The universality of Ginibre's correlations then implies universality of the cubic law of the eigenvalue repulsion, which can also be verified directly for complex ensembles with known jpdf \cite{Oas97}. Retaining all terms leads to a Fredholm determinant, see Chap. 4, giving access to various gap probabilities beyond the Ginibre ensemble \cite{Ake09a}.

\section{Random contractions}

Let $U$ be a unitary matrix taken at random from the unitary group ${\rm U}(M+L)$. Denote by $J$ its top left corner of size $M\times M$. The matrix $J$ is a random contraction: with probability one, it has all of its eigenvalues inside the unit disk $|z| < 1$. The jpdf of eigenvalues of $J$ was computed in \cite{Zyc00} and is given by (\ref{eq-b:5}) with
$w_L(z)=(1-|z|^2)^{L-1}\Theta (1-|z|)$. Since $w_L(z)$ is rotation invariant, the associated orthogonal polynomials are again powers $p_m(z)=z^m$  and the normalization constants are easily computed in terms of the Beta function, $h_m=\pi B(m+1,L)$. Applying the orthogonal polynomials formalism, one obtains the correlation functions in the determinantal form (\ref{eq-b:8a}) with
\[
K(z_1,z_2)=\frac{L}{\pi}\,
(1-|z_1|^2)^{\frac{L-1}{2}} (1-|z_2|^2)^{\frac{L-1}{2}} \sum_{m=0}^{M-1}
 {L+m \choose m} \,
 (z_1z_2^*)^m\, .
\]
The truncated binomial series for $(1-x)^{-(L+1)}$ on the rhs
can be expressed in terms of the incomplete Beta function
$I_x(a,b)=\frac{1}{B(a,b)}\int_{0}^x t^{a-1} (1-t)^{b-1}\, \rmd t$
via the relation $I_x(M, L+1) = 1- (1-x)^{L+1}\sum_{m=0}^{M-1}
 {L+m \choose m}
 \, x^m
$. 
This leads to a useful representation
\begin{equation}\label{eq-b:13-1a}
K(z_1,z_2)=\frac{L}{\pi}\,
\frac{(1-|z_1|^2)^{\frac{L-1}{2}} (1-|z_2|^2)^{\frac{L-1}{2}}}{(1-z_1z_2^*)^{L+1}}\, \Big(1-I_{z_1z_2^*}(M,L+1)\Big)\, .
\end{equation}
There are several asymptotic regimes to be considered in the context of $M\times M$ truncations of random unitary matrices of an increasing dimension $M+L$\footnote{The integral $I_x(a,b)$ is convenient for finding the relevant limits by the Laplace method.
E.g., one finds that $I_x(a,b) \simeq \Theta ( x-\frac{\alpha}{1+\alpha})$ as $a,b\to\infty$. The remainder term here is exponentially small when $x$ is away the transitional point $x_0=\frac{\alpha}{(1+\alpha)}$. In the transitional region
$I_{x_0+\frac{t}{\sqrt{a}}}\,(a,b) \simeq 1-\frac{1}{2}\erfc \big(\frac{(1+\alpha)^{3/2}}{\sqrt{2}\, \alpha}\, t \big)$.
Another asymptotic relation of interest is $I_{1-\frac{y}{a}}(a,b) \simeq {\Gamma(b,y)}/{\Gamma (y)}$ which holds for positive $y$ in the limit $a\to\infty$, $b$ is fixed.
}. The simplest one is the limit when $M$ stays finite and $L\to\infty$. In this regime the jpdf becomes Gaussian and one immediately recovers the complex Ginibre ensemble. In fact one can allow $M$ to grow with $L$ and still recover the Ginibre ensemble provided that $M\ll L$; see \cite{Jia06} and references therein for bounds on the rate of growth of $M$ for the Gaussian Law to hold. In the opposite case when the size of the truncation increases at the same rate as the overall dimension $M+L$, one has two distinct regimes: (i) $M, L\to \infty$, $M/L=\alpha>0$, and  (ii) $M\to\infty $, $L$ is finite.

We start with (i) which is the limit of strong non-unitarity. In this limit the eigenvalues are distributed inside the disk $|z|^2 \le \frac{\alpha}{(1+\alpha)}$ with density \cite{Zyc00}
\[
R_1(z)=\frac{L}{\pi}\, \frac{1-I_{|z|^{2}}(M,L+1)}{(1-|z|^2)^{2}}\ \simeq\  \frac{M}{\pi \alpha}\, \frac{1}{(1-|z|^2)^2}\, \Theta \Big(\frac{\alpha}{1+\alpha}-|z|^2 \Big)
\]
in the bulk and $R_1\big(\sqrt{\frac{\alpha}{1+\alpha}} +\frac{x}{\sqrt{M}} \big)\simeq \frac{M}{2\pi}\, \frac{(1+\alpha)^2}{\alpha}\, \erfc \big(\sqrt{2} \, \frac{1+\alpha}{\sqrt{\alpha}}\, x \big) $ at the edge. Modulo a simple rescaling, the edge density is the same as in the complex Ginibre ensemble. It can also be seen that the average number of eigenvalues outside the boundary $\simeq \sqrt{M(1+\alpha)/(2\pi)}$.

One can also find the eigenvalue correlation functions in the bulk. At the origin this task is especially simple. Scaling $z$ by $\sqrt{\pi L}$, which is the mean distance between the eigenvalues at the origin, and noticing that for $u$ in a bounded region in the complex plane
$\sum_{m=0}^{M-1} {L+m \choose m}\, \big(\frac{u}{L}\big)^m \simeq \rme^{u} $ in the strong non-unitarity limit, one concludes that the rescaled correlation functions are given by Ginibre's expression (\ref{eq-b:11a}). Although the eigenvalue distribution now is not homogeneous, after appropriate rescaling (\ref{eq-b:11a}) describes the eigenvalue correlations at any point in the bulk \cite{Zyc00}.

The gap probability at the origin, $H(s;0)$, can be computed by exploiting the rotational symmetry, in the same way as for the Ginibre ensemble, $H(s;0)=\prod_{m=1}^{M-1} \big(1- I_{s^2}(m+1,L) \big)$. After appropriate rescaling one obtains the same expression as in the Ginibre ensemble (cf.\ (\ref{eq-b:13})), $H\left(\frac{s}{\sqrt{L}}; 0\right) \simeq \prod_{m=1}^{\infty} {\Gamma(m+1, s^2)}/{\Gamma (m+1)}$ in the limit $M,L\to\infty$, ${M}/{L}=\alpha$, leading to the cubic law of of eigenvalue repulsion in the bulk.

Now, consider the limit of weak non-unitarity $M\to\infty $, $L$ is finite. In this limit the eigenvalues of $J$ lie close to the unit circle, with the magnitude of the typical deviation being of the order $1/M$.  Scaling $z$ accordingly, one finds the eigenvalue density in the limit of weak non-unitarity \cite{Zyc00}:
\[
R_1\Big(1-\frac{y}{M}\, \rme^{\rmi \phi}\Big)\ \simeq\  \frac{M^2}{\pi} \frac{(2y)^{L-1}}{(L-1)!} \int_0^1 \rme^{-2yt} t^L\, \rmd t\, ,\quad \mbox{$M\to\infty$ and $L$ is fixed}.
\]
Setting
$
z_j\!=\!\left(1-\frac{y_j}{M}\right)\, \rme^{\rmi \varphi_0+\rmi \frac{\varphi_j}{M}}
$, one finds the correlations in this limit:
\[
R_n(z_1,\ldots,z_n)\simeq \left(\frac{M^2}{\pi}\right)^{\!\!n} \prod_{j=1}^n  \frac{(2y_j)^{L-1}}{(L-1)!}\ \det \left(\int_0^1 \rme^{-(y_i+y_j+\rmi(\varphi_i-\varphi_j))t}\, t^L\, \rmd t\!  \right)_{i,j=1}^n\, .
\]
Interestingly, a different random matrix ensemble, $J=H+\rmi \gamma W$, leads to the same form of the correlation functions (after appropriate rescaling) as on the rhs above \cite{Fyo99}. Here $H$ is drawn from the GUE, $\gamma >0$ and $W$ is a diagonal matrix with $L$ 1's and $M$ zeros on the diagonal, with $M\gg 1$ and finite $L$. The above equation is a particular case of a universal formula describing correlations in more general ensembles of random contractions and non-Hermitian finite rank deviations from the GUE, see \cite{Fyo03} and Chap. 18 for discussion and results.

\section{Complex elliptic ensemble} \label{sec4}

Let $H_{1}$, $H_{2}$ be two independent samples from distribution $\rmd \mu (H) = \rme^{-\tr H^2} |DH|$ on the space of Hermitian $N\times N$ matrices. Then
$
J=\sqrt{1+\tau}H_1+i\sqrt{1-\tau} H_2
$
is a random matrix ensemble interpolating between the (circular) Ginibre ensemble ($\tau=0$) and the GUE ($\tau=1$). We only consider the interval $0\le \tau \le 1$. The matrix $J$ is complex Gaussian,
\begin{eqnarray}
\label{eq-b:18} \rmd \mu (J) &\propto& \exp \Big\{ -\frac{1}{1-\tau^2}\tr \left[JJ^{\dagger}  + \frac{\tau }{2} (J^2+ {J^{\dagger}}^2)\right]\Big\}\ |\rmD J|\, .
\end{eqnarray}
The jpdf, $P(z_1, \ldots, z_N)$, can be obtained by bringing $J$ to triangular form, as in Section \ref{sec2}. The resulting expression is given by (\ref{eq-b:5}) with weight function
$w_{\tau}(z)=\frac{1}{\pi \sqrt{1-\tau^2}}\
\exp\{-\frac{|z|^2}{1-\tau^2}  + \frac{\tau (z^2 + {z^*}^2)}{2(1-\tau^2)}\}\, $.
The associated orthogonal polynomials are scaled Hermite polynomials \cite{DiF94}. Indeed, by making use of the integral representation
$H_n(z)=\frac{n!}{2\pi \rmi}\oint {\rme^{2zt-t^2}}{t^{-(n+1)}}\ \rmd t$,
for the Hermite polynomials $H_n(z)$ one can easily verify that
\begin{equation}\label{eq-b:21}
\int_{\C} H_n\Big(\!\frac{z}{\sqrt{2\tau}}\Big)H_m\Big(\!\frac{z^*}{\sqrt{2\tau}}\Big) w_{\tau}(z)\ \rmd^2 z = \delta_{m,n}\, n! \Big(\frac{2}{\tau}\Big)^{n}\, .
\end{equation}
The monic polynomials $p_n(z)$ are easily found, $p_n(z)=(\tau/2)^{n/2}  H_n(z/\sqrt{2z})$, $h_n = n!$, and on applying the orthogonal polynomial formalism, one obtains the correlation functions in the determinantal form (\ref{eq-b:8a}) with
\begin{equation}\label{eq-b:22}
K(z_1,z_2)=w^{1/2}_{\tau}(z_1)w^{1/2}_{\tau}(z_2^*)\,
\sum_{n=0}^{N-1} \frac{\tau^n}{2^n n!}\, H_n\Big(\!\frac{z_1}{\sqrt{2\tau}}\Big) H_n\Big(\!\frac{z_2^*}{\sqrt{2\tau}}\Big)\, .
\end{equation}
The sum on the right is the truncated exponential series in Mehler's formula
\[
\sum_{n=0}^{\infty} \frac{\tau^n}{2^n n!} H_n\Big(\! \frac{z_1}{\sqrt{2\tau}}\Big) H_n\Big(\!\frac{z_2^*}{\sqrt{2\tau}}\Big)= \frac{1}{\sqrt{1-\tau^2}} \ \exp\left\{\frac{z_1z_2^*}{1-\tau^2} - \frac{\tau (z_1^2 + {z_2^*}^2)}{2(1-\tau^2)}\right\}.
\]
A quick comparison of (\ref{eq-b:22}) with Mehler's formula convinces that the density $K(z,z)$ is constant in the limit $N\to\infty$.  More care is needed to determine the boundary of the eigenvalue support. To this end, another integral for the Hermite polynomials comes in handy,
$H_n(z)={(\pm 2\rmi)^n} \int_{-\infty}^{+\infty} \rme^{-(t \pm \rmi z)^2} t^n\ \rmd t/{\sqrt{\pi}}$.
On substituting this into (\ref{eq-b:22}) and after a simple change of variables, one writes the kernel in a form suitable for asymptotic analysis
\begin{eqnarray}\label{eq-b:23a}
\hspace*{-8ex}K(z_1,z_2) &=& \frac{1}{\pi (1-\tau^2)} \exp \Big\{-\frac{|z_1|^2+|z_2|^2 - 2 z_1z_2^*}{2(1-\tau^2)}\Big\}\times \\
\label{eq-b:23b}
&& \int_{-\infty}^{+\infty}\!\! \rmd u \int_{-\infty}^{+\infty} \!\! \rmd v \ f^{-}(u) f^{+} (v)\  \frac{\Gamma (N, N \tau (u^2-v^2))}{\Gamma (N)}\, ,
\end{eqnarray}
with
$f^{\pm} (q)= \sqrt{\frac{N(1\pm \tau)}{\pi}}\, \exp\Big\{-\Big(q\sqrt{N(1\pm \tau)} - \frac{\rmi (z_1\pm z_2^*)}{2 \sqrt{\tau(1\pm \tau)}}\Big)^2\Big\}$.

For $N$ large with $1-\tau>0$ uniformly in $N$ and $x,y=O(\sqrt{N})$, the functions $f^{\pm}$ can formally be replaced by delta-functions, $f^{+} (v) = \delta \big(v-\frac{\rmi x}{(1+\tau)\sqrt{N\tau}}\big)$ and $f^{-} (u)=\delta \big(u+\frac{y}{(1-\tau)\sqrt{N\tau}}\big)$. Such a replacement can be justified by deforming the $v$-integral into the complex plane to pick up the sharp peak of $f^{+}(v)$ along the imaginary axis. This gives a simpler expression for the eigenvalue density
$R_1(z)\simeq (\pi (1-\tau^2))^{-1}\, \Gamma \big(N, \frac{x^2}{(1+\tau)^2} + \frac{y^2}{(1-\tau)^2}\big)/\Gamma (N)$, cf (\ref{eq-b:11bulk}).
Hence, for large $N$ the eigenvalues density is ${1}/{\pi (1-\tau^2)}$ inside the ellipse with half-axes $\sqrt{N}(1\pm \tau)$ along the $x$ and $y$ directions, in agreement with Girko's elliptic law \cite{Gir86}. It falls to zero exponentially fast when one transverses the boundary of the ellipse.

For $z_{1}$, $z_{2}$ inside the ellipse and such that $|z_1-z_2|=O(1)$ the integral in (\ref{eq-b:23b}) converges to 1 as $N\to\infty$, and after a trivial rescaling  one obtains the same expression (\ref{eq-b:11a}) for the correlations as in the circular case (at the origin this readily follows from Mehler's formula). 

The limit $N\to\infty$, $1-\tau>0$ is the limit of strong non-Hermiticity. There is another important limit, $N\to\infty$ and $1-\tau=\alpha^2/N$, the so-called limit of weak non-Hermiticity \cite{Fyo97} describing the cross-over from Wigner-Dyson to Ginibre eigenvalue statistics. In this limit the eigenvalues of $J$ can be thought of as those of $\sqrt{2}H_1$ (the Hermitian part of $J$) displaced from the real axis by the ``perturbation'' term $\rmi \frac{\alpha}{\sqrt{N}}H_2$ (the skew-Hermitian part of $J$). The eigenvalues $x_j$ of $\sqrt{2}H_1$ fill the interval $(-2\sqrt{N}, 2\sqrt{N})$ with density $\rho_{sc}(x) =\frac{1}{\pi} \sqrt{N-x^2/4}$ (Wigner's Semicircle Law). A simple perturbation theory calculation \cite{Fyo97} gives the density of eigenvalues $z=x+\rmi y$ of $J$ in the factorized form
$\rho_{sc}(x)\rho(y)$,
with the Gaussian distribution  $\rho(y)=\sqrt{{N}/{2\pi \alpha^2}}\, \exp(-{N y^2}/{2\alpha^2})$ of the displacements. For such a calculation to be well defined, the width of this distribution should be much smaller than the (mean) eigenvalue spacing $1/\rho_{sc}(x) $ of the unperturbed eigenvalues, making it natural introducing the control parameter
$a(x)= {\alpha \rho_{sc}(x)}/{\sqrt{N}}$.
Formulas (\ref{eq-b:23a})--(\ref{eq-b:23b}) make it possible to go beyond the perturbation theory and compute the eigenvalue density and correlations in the limit of weak non-Hermiticity exactly. In this limit the function $f^{+}(v)$ is singular and the same as in the limit of strong non-Hermiticity, however the function $f^{-}(u)$ is not. Let $z_j=x+\zeta_j/\rho_{sc}(x)$. Then, to the leading order,
$f^{-}(u) \simeq {\pi}^{-1/2} \alpha \exp {\big(u\alpha -\frac{\rmi (\zeta_1-\zeta_2^*)}{2\alpha \rho_{sc}(x)}\big)^2}$ and the integral in (\ref{eq-b:23b})
$\simeq 2\, \sqrt{\pi}\, a\, \exp\{\frac{(\zeta_1-\zeta_2^*)^2}{4a^2}\}\int_0^1 \rme^{-\pi^2a^2 u^2 } \cos (\pi u (\zeta_1-\zeta_2^*))\ \rmd u$,
where $a\equiv a(x)$ is the control parameter defined above.
One then obtains the scaled correlation functions $\tilde R_n(\zeta_1, \ldots, \zeta_n) = ({\rho_{sc}^2(x)})^{-n} R_n (z_1, \ldots, z_n)$ in the form \cite{Fyo97}:
\begin{equation}\label{eq-b:26}
\tilde R_n(\zeta_1, \ldots, \zeta_n) \simeq \left( \frac{1}{\sqrt{\pi} a}\right)^n\!\! \rme^{-\frac{1}{a^2} \sum_{j} (\im \zeta_j)^2 }\! \det \left( \! \int_0^1\! \rme^{-\pi^2a^2 u^2 } \cos (\pi u (\zeta_i-\zeta_j^*))\, \rmd u \!\right) .
\end{equation}
It is easy to check that $\tilde R_n$ interpolates between the Wigner-Dyson correlations ($a\to 0$) and Ginibre's ($a\to \infty$), and the above equation allows one to study the cross-over from one set of eigenvalue statistics to the other, see \cite{Fyo98}. The eigenvalue density is
\begin{equation}\label{eq-b:27}
\tilde R_1(\zeta)\simeq \frac{1}{\sqrt{\pi} a}\  \rme^{-\frac{(\im \zeta)^2}{a^2}  } \int_0^1 \rme^{-\pi^2a^2 u^2 } \cosh (2\pi u \im \zeta )\ \rmd u\, ,
\end{equation}
correcting the perturbation theory result.  Interestingly, the distribution of the scaled imaginary parts of eigenvalues as given in (\ref{eq-b:27})
appears to be universal. A supersymmetry calculation shows \cite{Fyo98} that it does not depend on the details of the Hermitian and skew-Hermitian parts of $J$ and extends to non-invariant matrix distributions (like the semicircular and circular laws). It is an open challenging problem to find a proof of this universality satisfying the rigor of pure mathematics. Staying within the class of invariant distributions, the eigenvalue density and correlations appear to be also universal: it was argued in \cite{Ake03} that the weakly non-Hermitian limit in the ensemble (\ref{eq-b:1a}) at the origin is also described by (\ref{eq-b:26}).

In conclusion, we would like to mention briefly two topics related to the complex elliptic ensemble. One is the recent studies of edge scaling limits \cite{Gar02,Ben08}. By scaling $\tau$ with $N$ so that $1-\tau=\alpha/N^{1/3}$ \cite{Ben08} one gets access to the crossover from Airy ($\alpha \ll 1$, GUE) to Poisson edge statistics ($\alpha \gg 1$, Ginibre). And the other  is the chiral extension
$J=
\left(
\begin{smallmatrix}
     0 & \rmi A + \mu B \\
     \rmi A^{\dagger} + \mu B^{\dagger} & 0
\end{smallmatrix}
\right)
$
of the complex elliptic ensemble. Here $A$ and $B$ are two independent samples from the Gaussian measure with density $\rme^{-\tr A^{\dagger}A}$ on the space of complex $(N+\nu)\times N$ matrices, $\nu\ge 0$, $0\le \mu\le 1$. This ensemble can be studied along the same lines as (\ref{eq-b:18}). The computation of the jpdf of eigenvalues \cite{Osb04} is though more involved and leads to (\ref{eq-b:5}) with weight function $w(z) = |z|^{2\nu +2}e^{-\frac{1-\mu^2}{4\mu^2}\,(z^2+{z^*}^2)}\, K_{\nu}(\frac{1+\mu^2}{2\mu^2}|z|^2)$ where $K_{\nu}$ is a modified Bessel function with the associate orthogonal polynomials being scaled Laguerre polynomials \cite{Osb04,Ake05}. At $\mu=0$ we recover the Wishart (Laguerre) ensemble of Hermitian matrices and the corresponding weakly non-Hermitian limit $N\to\infty$ and $a=N\mu^2=O(1)$ describes its neighborhood. The eigenvalue correlations in this limit were computed in \cite{Osb04}, with the answer being somewhat different from (\ref{eq-b:26}). This is not surprising given that the two ensembles belong to different symmetry classes. By letting $a\to\infty$ one obtains the correlations in the chiral ensemble in the regime of strong non-Hermiticity, see, e.g., \cite{Ake05a}.

\section{Real and quaternion-real Ginibre ensembles}

{\bf Measure, change of variables} \quad Restricting to even dimensions $N$, we will treat the real and quaternion real (qu-r) ensembles in a unifying way. Consider the normalized measure
\begin{equation}\label{muJ}
    {\rm d}\mu(J)=\exp( -\frac{1}{2}\tr J J^{\dagger} )\,  |{\rm D} J |
\end{equation}
on the space of $N\times N$ matrices subject to the constraints
\begin{equation}\label{JT}
\begin{array}{ll}
J^{T}=J^{\dagger} & \text{in the real case}\\[1ex]
Z J^{T} Z^{T}=J^{\dagger} & \text{in the quaternion-real (qu-r) case}\, ,
\end{array}
\end{equation}
where $Z= \bigoplus_{j=1}^{N/2}{\ 0\ 1\choose-1\  0}$ is the symplectic unit. Here ${\rm D} J =\prod_{i,j=1}^N ({\rm d}J_{ij}/\sqrt{2\pi})$ is the (exterior) product of the one-forms in matrix entries and $|\rmD J|$ is the corresponding Cartesian volume element.

As follows from (\ref{JT}), both real and qu-r matrices have non-real eigenvalues occurring in pairs $z$ and $z^*$, and any real eigenvalue of a qu-r matrix has multiplicity $\ge 2$. Hence, the probability for a qu-r matrix drawn from the distribution (\ref{muJ}) to have a real eigenvalue is zero.

As with complex matrices, the jpdf of eigenvalues can be obtained from a Schur decomposition. Because of the symmetries, it is  convenient to work with matrices partitioned into $2\times 2$ blocks. Ignoring multiple eigenvalues, we can bring $J$ to block-triangular form, $J= U(\Delta + \Lambda)U^{-1}$ where $\Lambda$ is block-diagonal and $\Delta$ has nonzero blocks only above $\Lambda$. The matrix $U$ is orthogonal in the real case and unitary symplectic in the qu-r case and can be restricted to the space of right cosets, ${\rm O}(N)/{\rm O}(2)^{N/2}$ and ${\rm USp}(N)/{\rm USp}(2)^{N/2}$ respectively.

The variations in $J$ are related to those in $U, \Lambda, \Delta$ by
\begin{equation}\label{b:s}
    U^{-1} {\rm d} J\, U =  {\rm d}\Lambda + {\rm d}\Delta + [U^{-1}{\rm d}U, \Delta] + [U^{-1}{\rm d}U, \Lambda]\, ,
\end{equation}
with $[A,B]=AB-BA$. The matrix $U^{-1}{\rm d}U$ is skew-symmetric in the real case, $(U^{-1}{\rm d}U)^{T}=-(U^{-1}{\rm d}U)$, and anti self-dual in the qu-r case, $Z(U^{-1}{\rm d}U)^{T}Z^{T} =-(U^{-1}{\rm d}U)$.

To find the Jacobian associated with the change of variables from $J$ to $U,\Lambda, \Delta$, we multiply entries of the matrix on the rhs in (\ref{b:s}) using the calculus of alternating differential forms. The matrices ${\rm d}\Lambda $ and ${\rm d}\Delta$ yield the volume forms for $\Lambda$ and $\Delta$, respectively. Due to the triangular structure, the matrix $[U^{-1}{\rm d}U, \Delta]$ does not contribute, and the matrix $[U^{-1}{\rm d}U, \Lambda]$ yields the coset volume form times a factor depending on the eigenvalues $\lambda_j$ of $\Lambda$,
\[
{\prod}'(U^{-1}{\rm d}U\Lambda- \Lambda U^{-1}{\rm d}U)_{ij}={\prod}' (U^{-1}{\rm d}U)_{ij}(\lambda_j-\lambda_i)\, ,
\]
with the dashed product running over non-zero entries in the lower triangle of $U^{-1}{\rm d}U$. On gathering all terms, we arrive at
\begin{equation}\label{muJ2}
     {\rm d}\mu(J)=\rme^{ -\frac{1}{2}\tr (\Delta \Delta^{\dagger}+ \Lambda \Lambda^{\dagger})}\, |{\rm D} \Delta |\, |{\rm D}\Lambda|\, |{\prod}'(U^{-1}{\rm d}U)_{ij}(\lambda_j-\lambda_i)/\sqrt{2\pi}|
\end{equation}
with ${\rm D}\Lambda=\prod\ ({\rm d}\Lambda_{ij}/\sqrt{2\pi})$ and ${\rm D}\Delta=\prod\ ({\rm d}\Delta_{ij}/\sqrt{2\pi})$, the products running over nonzero entries. The jpdf of eigenvalues follows from (\ref{muJ2}) on integrating out all auxiliary variables. Since the density function does not depend on $U$, the $U$-integral gives the volume of the coset space. The integral in $\Delta$ is Gaussian, and this variable can be integrated out with ease as well. Thus we are left with the problem of integrating over the $N/2$ $(2\times2)$-blocks appearing in $\Lambda$ keeping the eigenvalues in each block fixed. These are precisely the eigenvalues of the matrix $J$ and our problem reduces to $2\times2$ matrices.

{\bf Dimension $N=2$ and the jpdf} \quad
The generic form of a qu-r $2\times 2$ matrix is
$ J= \left(
\begin{smallmatrix}
     a & b\\
    -b^* & a^*
\end{smallmatrix}
\right)
$
with $a$ and $b$ complex. The eigenvalues are complex conjugate, $\lambda_{1,2} = {\rm Re}\ a \pm \rmi \sqrt{({\rm Im}\ a)^2 + |b|^2}$, and we choose $\im \lambda_1= -\im \lambda_2 > 0$. Thus, $\int \rmd \mu(J)\,  \delta^2(\lambda_1- \re a - \rmi  \sqrt{(\im a)^2 + |b|^2}) =\frac{|\lambda_1-\lambda_2|}{\pi}\, f^{q}(\lambda_1)^2$ with $f^q(\lambda)$ as in (\ref{frq}). The rhs is normalized wrt integrating over $\lambda_1$ in the half-plane $\im \lambda_1 >0$.

Turning to real $2\times2$ matrices
$
    J= \left( \begin{smallmatrix} a & b\\
    c & d
    \end{smallmatrix} \right)
$
with $a, b, c, d$ real, the eigenvalues are $\lambda_{1,2} = \frac{1}{2}(a+d \pm  \sqrt{(a-b)^2 - 4 bc})$, and we now have two possibilities (i) $\lambda_1=\cc{\lambda_2}$ (choosing $\im \lambda_1= - \im  \lambda_2>0$), or (ii) both $\lambda_1$ and ${\lambda_2}$ are real (choosing $\lambda_1>\lambda_2$). In the first case we obtain \cite{Leh91,Som08}
$\int \rmd \mu(J)\,  \delta^2(\lambda_1 - \frac{1}{2}[a+d - \rmi  \sqrt{4bc-(a-b)^2}])
= \frac{|\lambda_1-\lambda_2|}{\sqrt{2\pi}} f^r(\lambda_1)^2$
with $f^r(\lambda)$ as in (\ref{frq}).
In the 2nd case ($\lambda_{1,2}$ real) the corresponding average over the product of two $\delta$-constraints results in   $\frac{|\lambda_1-\lambda_2|}{2\sqrt{2\pi}}\, f^r(\lambda_1)\,f^r(\lambda_2)$ \cite{Som08}.

Consequently, the joint distribution of eigenvalues for both ensembles real and qu-r can be written in the form \cite{Gin65,Meh04,Som08}
\begin{equation}
\label{jpdfR}
      {\rm d}\mu(\lambda_1,\lambda_2,...,\lambda_N)=  C_N \prod_{1\le i < j \le N}\!\! (\lambda_i-\lambda_j)\, \prod_{i=1}^N f(\lambda_i)\ \rmd \lambda_1 \rmd \lambda_2 \cdots \rmd \lambda_N \, ,
\end{equation}
where
\begin{equation}\label{frq}
\hspace*{-2ex}
f(\lambda)^2 =
\begin{cases}
f^r(\lambda)^2= {\erfc \Big(\frac{|\lambda-\lambda^*|}{\sqrt{2}}\Big)}\, \rme^{-\frac{1}{2}(\lambda^2+{\lambda^*}^{2})} & \text{in the real case}\\[1.5ex]
f^q(\lambda)^2= {|\lambda-\lambda^*|}\, \rme^{- |\lambda|^2} & \text{in the qu-r case},
\end{cases}
\end{equation}
with $f(\lambda)=f(\lambda^*)\ge 0$ in both cases. We have to put the eigenvalues in such an order that ${\rm d}\mu\ge0$.
Thus we consider each case with $m$ complex conjugate pairs of eigenvalues ($m=0,1,2,...,N/2$) separately, arranging the $\lambda_i$'s in (\ref{jpdfR}) in the following order: for $m=0$ (all real) $\lambda_1>\lambda_2>...>\lambda_N$, for $m=1$ (one complex conjugate pair $\lambda_1= \cc{\lambda_2}$) $\im \lambda_1>\im \lambda_2,\ \lambda_3>\lambda_4>...>\lambda_N$, for $m=2$ (two complex conjugate pairs $\lambda_1= \cc{\lambda_2}$ and $\lambda_3= \cc{\lambda_4}$) $\im \lambda_1>\im \lambda_2, \re \lambda_2>\re \lambda_3, \ \im \lambda_3>\im \lambda_4,\ \lambda_5>...> \lambda_N$, etc.. Summing and integrating over all cases and ranges yields total probability $1$. In the qu-r case we only have $m=N/2$ complex conjugate pairs, and in the real case the probabilities of finding $m$ complex conjugate pairs have been calculated in \cite{Kan05b,For07}.

The jpdf (\ref{jpdfR}) shows immediately the repulsion behavior of the eigenvalues. Due to the Vandermonde it contains for two eigenvalues in the upper halfplane the factor $|\lambda_1-\lambda_2|^2$ which means cubic repulsion in the distance, the additional power coming from the two-dimensional volume element  \cite{Gro88}. For two eigenvalues exactly on the real axis one has the factor $|\lambda_1-\lambda_2| $ and thus linear repulsion like in GOE. This applies only in the real case. In this case due to the factor $|\lambda_1-\lambda_1^*|$ there is linear repulsion of complex
eigenvalues with distance from the real axis, while in the qu-r case there is quadratic repulsion from the real axis due to an additional factor
$|\lambda_1-\lambda_1^*|$ coming from $f^q(\lambda_1)^2$.

{\bf Correlation functions as Pfaffians} \quad It is convenient to abandon the $\lambda_j$'s in favour of the real $2$-dimensional vectors $z_j=\lambda_j$ with $z_j=x_j+\rmi y_j$ and the two dimensional volume element ${\rm d}^2z={\rm d}x\, {\rm d}y$ (in contrast, in (\ref{jpdfR}) ${\rm d}\lambda{\rm d}\lambda^* =-2\rmi {\rm d}x\, {\rm d}y$). Equation (\ref{jpdfR}) defines the symmetrized jpdf of eigenvalues $P(z_1, \ldots, z_N)$ in the obvious way (however, care must be taken to account for complex conjugate pairs and the ordering of eigenvalues before symmetrizing the density). $P(z_1, \ldots, z_N)$  is a formal density, as it contains delta-functions accounting for complex conjugate pairs and also for eigenvalues on the real line.

The eigenvalue correlation functions (\ref{eq-b:4a}) (more precisely they are measures, since they contain delta-function contributions) can be obtained by functional derivatives from the generating functional \cite{Tra98,Kan02,Som08}
\begin{equation}\label{GF}
    Z[g]=\int {\rm d}^2z_1 \ldots \int {\rm d}^2z_N \ g(z_1)\ldots g(z_N)\,  P(z_1,z_2,...,z_N)
\end{equation}
as
$
R_n(z_1,z_2,\ldots,z_n)=\frac{\delta^n}{\delta g(z_1)...\delta g(z_n)}\, Z[g] \, \big|_{g(z)\equiv 1}.
$
They can also be obtained by integration from the symmetrized jpdf via (\ref{eq-b:4a}). This implies a simple integration theorem for the full correlations $R_n$ (an Pfaffian analogue of Mehta's 'integrating out' lemma, see (\ref{eq-b:8a})). Restricting to the smooth (in the upper half-plane) and singular (on the real line) parts of the $R_n$ leads to a more sophisticated integration theorem \cite{Kan05b} connecting these quantities recursively. We find it more convenient to work with the generating function $Z[g]$ which can be calculated in the form of a Pfaffian with the help of the Grassmann integral representation of the Vandermonde determinant \cite{Som08}
$\prod_{1\le i<j\le N} (z_i-z_j) = \int{\rm d}\chi_1\ldots{\rm d}\chi_N\prod_{i=1}^N(\sum_{k=1}^Nz_i^{k-1}\chi_k)$
with $\chi_k\chi_l=-\chi_l\chi_k$. Since the integrand $g(z_1)\ldots g(z_N)$ in (\ref{GF}) is symmetric in the $z_i$'s, one can integrate it against the joint distribution of ordered eigenvalues as in (\ref{jpdfR}) in order to obtain $Z[g]$. This helps to avoid the absolute values in the Vandermonde determinant, making it possible to apply the Grassmann integral above. Recalling the Grassmann integral representation $\Pfaff (A_{kl})=\int{\rm d}\chi_1\ldots {\rm d}\chi_N\exp(-\frac{1}{2}\sum_{kl}\chi_kA_{kl}\chi_l)
$ for the Pfaffian of an antisymmetric matrix $(A_{kl})$
one finds immediately
\begin{equation}\label{Z[g]}
Z[g]=C_N\, \Pfaff  ({ \tilde A}_{kl}),
\end{equation}
with ${ \tilde A}_{kl} =  \int\int\! {\rm d}^2z_1{\rm d}^2z_2\, {\cal F}(z_1, z_2)\, z_1^{k-1} z_2^{l-1}\, g(z_1)g(z_2)$ and
\begin{equation}\label{calFR}
{\cal F} (z_1, z_2 ) = f(z_1) f(z_2) (\, 2\rmi\, \delta^2(z_1-\cc{z_2}) \sgn(y_1) + \delta(y_1) \delta(y_2) \sgn(x_2-x_1)\, )\, .
\end{equation}
This simple form of $\cal F$ is a consequence of the fact that in (\ref{jpdfR}) the normalization constant does not depend on the chosen number $m$ of complex conjugate pairs. The first and second terms on the rhs account for eigenvalues coming in complex conjugate pairs and eigenvalues on the real axis, respectively. The second term vanishes in the qu-r case since $f^q=0$ on the real axis.

Putting $g(z)=1+u(z)$ and expanding $Z[1+u]$ in powers of $u$ one again obtains a series of Pfaffians \cite{Som08}. This method goes back to Mehta's alternate variables \cite{Meh04} and Tracy and Widom's paper \cite{Tra98}. With $\tilde A_{kl}\big \vert_{g\equiv 1}=A_{kl}$, $A=(A_{kl})$ and defining the kernel
\begin{equation}\label{calK}
{\cal K}_N (z_1, z_2 ) =   \sum_{k=1}^N \sum_{l=1}^N  A_{kl}^{-1} z_1^{k-1} z_2^{l-1}
\end{equation}
the $n$-point densities are given by
\begin{equation}\label{Rn2}
    R_n(z_1, \ldots , z_n) = {\rm Pfaff} (Q_{kl}) \quad \mbox{with}\ Q_{kl} = \left( \begin{array}{cc} K_{kl} & G_{kl}\\
    -G_{lk} & W_{kl}
    \end{array} \right)\ .
\end{equation}
This means the Pfaffian of the $2n\times 2n$ matrix built of the $n^2$ quaternions ($2\times 2$ matrices) $Q_{kl}$, $k,l=1,2,\ldots, n$ with
entries  $G_{kl} = \int {\rm d}^2z \, {\cal K}_N(z_k,z){\cal F} (z,z_l)$, $K_{kl} ={\cal K}_N(z_k,z_l)$, and $W_{kl} = \int {\rm d}^2z  \int {\rm d}^2z'\, {\cal F} (z_k,z){\cal K}_N(z,z'){\cal F} (z',z_l)- {\cal F} (z_k,z_l) $ and symmetries $K_{kl}=-K_{lk}$, $W_{kl}=-W_{lk}$.
With the above expressions  one finds, e.g., the one-point density:
\begin{equation}\label{R1}
    R_1(z_1) = \int {\rm d}^2z_2 \, {\cal F} (z_1,z_2) {\cal K}_N(z_2,z_1) = R_1^C(z) + \delta(y)R_1^R(x)
\end{equation}
and the two-point density, see Eq. (21) in \cite{Som07}.
Note that ${\cal F}(z_1,z_2)$ is composed of two parts. Correspondingly, $R_1 (z)$ contains two parts, a smooth part $R_1^C(z)$ which describes the density of complex eigenvalues and a singular part $R_1^R(x)$ that describes the density of real eigenvalues.
In the qu-r case $R_1^R(x)=0$ and only the complex part remains. Restricting oneself to points $z_1,z_2,\ldots,z_n$ in the upper half-plane no terms containing delta functions like $\delta^2(z_i-z_j^*)$ appear in the correlation functions and (\ref{Rn2}) gives the correlations directly in terms of Pfaffians involving the kernel ${\cal K}_N$. Also, the real correlations can be obtained by considering only the terms concentrated on the real axis \cite{Som07,For07}.   These however  involve still some real integrations \cite{Som07}.

{\bf Kernel, characteristic polynomials}\quad In order to use (\ref{Rn2}) one needs to know the kernel ${\cal K}_N(z_1,z_2)$ which we now are going to find. It follows from (\ref{R1}) and (\ref{calFR}) that $R^C(z)=2f(z)f(z^*)|{\cal K}_N(z,z^*)|$. On the other hand one can find $R_1^C(z)$ (and $R_1(z)$) directly from the jpdf ( \ref{jpdfR}) by integrating out $N-1$  variables. Since the eigenvalues are real or come in complex conjugate pairs this leads to $
R_1^C(z)\propto |z- z^*|\langle\det(J-z)\ \det(J- z^*)\rangle_{N-2},
$  where $\langle \ldots \rangle_{N-2}$ means averaging over the ensemble (\ref{muJ}) in $N-2$ dimensions \cite{Ede97}. On comparing the two expressions for $R_1^C(z)$ one arrives at the important relation:\footnote{One can also express the kernel and/or correlation functions via averages of the characteristic polynomials in the qu-r and complex Ginibre ensembles  and in some ensembles beyond Gaussian. The averages like on the rhs in (\ref{CP}) can be computed in a variety of ways and this gives and an alternative way of calculating the eigenvalue correlation functions in the complex plane in a variety of random matrix ensembles, see \cite{Ede94,Fyo99,Ake03,Ake07,Fyo07,Ake09}.}
\begin{equation}
\label{CP}
 {\cal K}_N(z, z^*) \propto (z- z^*) \langle\det(J-z)\, \det(J -z^*)\rangle_{N-2}\, .
       \end{equation}
Thus, we are left with the task of calculating
$\langle\det(J-u)\ \det(J-v)\rangle_{N}$.
Writing the determinants as Grassmann integrals, $\det(J-u)=\int {\rm D}\eta  \rme^{-\eta^{\dagger}(J-u)\eta}$,
\begin{equation} \label{Puv2}
 P_N(u,v) =\langle\det(J-u)\det(J-v)\rangle_{N}=\int {\rm D}\eta {\rm D}\zeta\  \rme^{u\eta^{\dagger}\eta +v\zeta^{\dagger}\zeta}\  \rme^{\frac{1}{2}\langle (\tr J (\eta \eta^{\dagger} + \zeta \zeta^{\dagger} )^2\rangle_N}
\end{equation}
where we have used the Gaussian property. The lowest cumulants of $J$ are $\langle J_{ij}\rangle_N=0$ and $\langle J_{ij} J_{kl}\rangle_N= \Delta_{ik}\Delta_{jl}$, where in the real case $\Delta_{ik}=\delta_{ik}$,  while in the qu-r case $\Delta_{ik}=Z_{ik}$ with $Z$ being  the symplectic unit.

In the real case, after a complex Hubbard-Stratonovich (HS) transformation,
\begin{equation}
 P_N(u,v)= \int \frac{{\rm d}^2a}{\pi}\ {\rm e}^{-|a|^2}\! \left(\int{\rm d}\eta^{*} {\rm d}\eta {\rm d}\zeta^{*} {\rm d}\zeta {\rm e}^{u \eta^{*} \eta + v \zeta^{*} \zeta + i a \eta ^{\dagger} \zeta^{\dagger}+ i \bar a \eta \zeta} \right)^N\, .  \label{Puv3}
\end{equation}
On evaluating the simple Grassmann integration, one arrives at
\begin{equation}\label{Puv4}
 P_N(u,v)= \int \frac{{\rm d}^2a}{\pi}\, {\rm e}^{-|a|^2} (uv+|a|^2)^N = N!\sum_{n=0}^N\frac{(uv)^n}{n!}\, .
\end{equation}
After restoring the normalization one finds the kernel
\begin{equation}
\label{10}
{\cal K}_N(z_1,z_2) = \frac{z_1 - z_2}{2 \sqrt{2 \pi}} \sum_{n=0}^{N-2} \frac{(z_1 z_2)^n}{n!} = \frac{z_1 - z_2}{2 \sqrt{2 \pi}}\ \rme^{z_1 z_2} \, \Gamma (N-1,\, z_1 z_2)\, .
\end{equation}
This gives immediately the density of complex eigenvalues \cite{Ede97}, cf (\ref{eq-b:11bulk}),
\begin{equation}
\label{11a}
R_1^C(z) = \frac{2 |y|}{\sqrt{2\pi}}\ \rme^{2y^2}\!\erfc(\sqrt{2}|y|)\,  \frac{\Gamma (N-1, |z|^2)}{\Gamma(N-1)}, \quad z=x+\rmi y\, .
\end{equation}
and, in view of (\ref{R1}), the density of real eigenvalues \cite{Ede94}
\begin{equation}
\label{11}
R_1^R(x)=\frac{\Gamma(N-1,x^2)}{\sqrt{2\pi} \Gamma(N-1)} + \frac{{\rm
e}^{-x^2/2} x^{2N-2}}{2^{N-1/2} \Gamma(N/2)} \gamma^*\Big(\frac{N-1}{2},
\frac{x^2}{2}\Big)\, ,
\end{equation}
with $\gamma^*(N,x)=x^{-N}(1-{\Gamma(N,x)}/{\Gamma(N)})$.
The rhs in (\ref{11a}) and (\ref{11}) is analytic in $N$ and these equations hold for odd $N$ as well \cite{Som08,For09a}.

One can easily analyze (\ref{11a}) and (\ref{11}) in the limit $N\to\infty$. For example, the average number $\langle n_R \rangle_N$ of real eigenvalues can be found by integration,  recovering the result by Edelmann, Kostlan and Shub \cite{Ede94}
\begin{equation}
\label{EKS}
\langle n_R \rangle_N = 1 + \frac{\sqrt{2}}{\pi} \int_{0}^{1} \frac{{\rm d}t \; t^{1/2} ( 1 - t^{N-1})}{(1-t)^{3/2} (1+t)} \simeq \sqrt{\frac{2N}{\pi}}
\end{equation}
and verifying the conjecture $\langle n_R \rangle_N \propto \sqrt{N}$ made in \cite{Leh91}. Interestingly, the variance of $n_R$ is also proportional to $\sqrt{N}$ \cite{For07}. For large $N$ and away from the real line ($|y|\gg 1$) the density of complex eigenvalues obeys the circular law (\ref{eq-b:11bulk}) with the same edge profile (\ref{eq-b:11edge}) as in the complex Ginibre ensemble. Inside the circle $|z|<\sqrt{N}$ and close to the real line ($y=O(1)$) $R_1^C(z) \simeq \sqrt{2/\pi}\, |y|\, \rme^{2y^2}\!\erfc(\sqrt{2}|y|)$. One also finds a constant density of real eigenvalues inside the same circle\cite{Ede94} $R_1^R (x) \simeq \frac{1}{\sqrt{2\pi}} \Theta (\sqrt{N} - |x|)$ which is consistent with the asymptotics in (\ref{EKS}). This constant density is in contrast to the Wigner Semicircle Law for Gaussian Hermitian or real symmetric matrices. In the transitional region around the end point $x=\sqrt{N}$ \cite{For07,Bor08}
\[
R_1^R(\sqrt{N} + u) \simeq \frac{1}{2\sqrt{2\pi}}\erfc(\sqrt{2}u) + \frac{1}{4\sqrt{\pi}}\, \exp(-u^2)\erfc (- u)\, .
\]
Although in the qu-r case  
${\cal K}_N(z_1,z_2)$
can be obtained as saddle point integral 
similarly to (\ref{Puv4}), it is 
convenient to derive it directly from the integral $A_{kl}$
\begin{equation}   \label{Akl}
A_{kl}=\int{\rm d}^2z\ 2(z-z^*){\rm e}^{-|z|^2}z^{k-1}{z^*}^{l-1}=2\pi(k!\, \delta_{k,l-1}-l!\, \delta_{l,k-1})\, .
\end{equation}
We see that in the qu-r case the matrix  $(A_{kl})$ has a simple tridiagonal structure, while in the real case its inverse is tridiagonal (as evident from (\ref{10})). Due to this structure we were able to find a beautiful formula for all Schur function averages (moments) in the real Ginibre ensemble \cite{Som09}. With this formula one can calculate the moments of symmetric functions in eigenvalues by making use of the Schur function expansion. A similar formula exists for the qu-r Ginibre ensemble \cite{For09b}.

Using the duplication formula for $\Gamma (z)$ and introducing  
$a_k=2^{k/2}\Gamma(k/2)$,
\[
A_{kl}=-\sqrt{{\pi}/{2}}\, a_{k+1}a_{l+1}\, \epsilon_{kl}\ \text{in the qu-r case}, \quad A_{kl}= a_k\,a_l\, \epsilon^{-1}_{kl}\ \text{in the real case,}
\]
where  $(\epsilon_{kl})=\tridiag(1,0,-1)$ is the antisymmetric tridiagonal matrix with 1's below the main diagonal and -1's above and its inverse (also antisymmetric) $\epsilon^{-1}_{kl}=1$ for $k$ odd and $j$ even in the upper triangle $k<j$ and the remaining entries being zero in that triangle. From these formulas and (\ref{calK}) one finds the kernel ${\cal K}_N(z_1,z_2)$ in the qu-r case thus recovering Mehta's result \cite{Meh04}
\[
{\cal K}_N(z_1,z_2)= - \frac{1}{\sqrt{2\pi}}\sum_{k=1}^N \sum_{l=1}^N
\frac{2^{-(k+l)/2}}{\Gamma((k+1)/2))\, \Gamma((l+1)/2)}\,
\epsilon^{-1}_{kl}\, z_1^{k-1} z_2^{l-1}\, .
\]
Also, the normalization constant $C_N$ in (\ref{jpdfR}) follows via (\ref{Z[g]})
\begin{equation}\label{cs}
\frac{1}{C_N}=
\begin{cases}
\left(\frac{\pi}{2}\right)^{N/4}\prod_{k=1}^N a_{k+1}=(2\pi)^{N/2}\ 1!\ 3!...(N-1)! & \text{in the qu-r case}\\
\prod_{k=1}^N a_k=(2\sqrt{2\pi})^{N/2}\ 0!\ 2!...(N-2)! & \text{in the real case.}
\end{cases}
\end{equation}
We see that the problem with all correlations and moments is to calculate the matrix $A_{kl}$ and its inverse $A^{-1}_{kl}$. In the qu-r case the calculation of $A_{kl}$ is simple and leads then to $A^{-1}_{kl}$. In the real case the calculation of $A^{-1}_{kl}$ via characteristic polynomials is simple and a direct calculation of $A_{kl}$ and  taking then the inverse is much more involved but possible with the help of the method of skew orthogonal polynomials wrt to the form ${\cal F}(z_1,z_2)$, as was demonstrated in \cite{For07,Bor08}. The method of skew-orthogonal polynomials can also be used to derive the correlation functions in the qu-r ensemble \cite{Kan02}. We will quote these results in connection with more general elliptic ensembles.

\section{Real and qu-r elliptic ensembles}
Consider now the two families of normalized measures
\begin{eqnarray}\label{mJt}
 {\rm d}\mu_{\tau}(J)&=& B_{\tau}\, \rme^{ -\frac{1}{2(1-\tau^2)}\tr( J J^{\dagger}-\frac{\tau}{2}(J^2+{J^{\dagger}}^2))}\  |{\rm D} J |
     \end{eqnarray}
on the space of $N\times N$ matrices subject to the symmetry constraints (\ref{JT}). The normalization constant $B_{\tau}$ is $ (1-\tau)^{-N(N-1)/2}(1+\tau)^{-N(N+1)/2}$ in the real case and  $(1+\tau)^{-N(N-1)/2}(1-\tau)^{-N(N+1)/2}$ in the qu-r case. When $\tau$ varies between $-1$ and $+1$ $ $  ${\rm d}\mu_{\tau}$ interpolates between anti-Hermitian (antisymmetric or antiselfdual, $\tau=-1$) and Hermitian (symmetric or selfdual, $\tau=+1$) ensembles.
On comparing to (\ref{muJ}), one concludes that the jpdf of eigenvalues in ensembles (\ref{mJt}) can simply be obtained by a rescaling of the term $\tr JJ^{\dagger}$ and otherwise multiplying with the factor $\exp(\lambda^2+{\lambda^*}^2)\tau/4(1-\tau^2))$, thus leading to the same expression (\ref{jpdfR}) with $f(\lambda)$ replaced by $f_{\tau}(\lambda)$, with
$f_{\tau}^2(\lambda)=f^2({\lambda}/{\sqrt{1-\tau^2}})\, \exp\{\frac{\tau}{2(1-\tau^2)}\, (\lambda+ {\lambda^*}^2)\}$
and a new normalization constant $C_{N,\tau}$. Correspondingly, the skew-symmetric form is given by (\ref{calFR}), again with $f(z)$ replaced by $f_{\tau}(z)$. Hereby  the matrices $A_{kl}$, $A^{-1}_{kl}$, the kernel ${\cal K}_N(z_1,z_2)$ and the correlations can be obtained.

{\bf Kernel in the elliptic case}\quad The kernel is again related to the average of two characteristic polynomials (\ref{CP})--(\ref{Puv2}), but now the second cumulants of $J$ are
$ \langle J_{ij} J_{kl}\rangle_N= \Delta_{ik}\Delta_{jl} + \tau\ \Delta_{il}\Delta_{jk}$
with $\Delta_{ik}$ as before.

In the real case one can evaluate the corresponding Grassmann integral by introducing two complex and two real HS transformations. This gives
\begin{equation}
P_N(u,v)= \int \frac{{\rm d}^2a}{\pi}\frac{{\rm d}^2b}{\pi}   {\rm e}^{-|a|^2-|b|^2 }\int\frac{{\rm d}x{\rm d}y}{2\pi}{\rm e}^{-(x^2+y^2)/2 } P_{r,\tau}^{N}   \label{Prtau}
\end{equation}
where $P_{r,\tau}$ is the Pfaffian of a $4$-dimensional antisymmetric matrix equivalent to a 4-fold Gaussian Grassmann integral with the value
$P_{r,\tau}=(u+\rmi x\sqrt{\tau})(v+\rmi y\sqrt{\tau}) +|a|^2 + \tau |b|^2$.
From (\ref{Prtau}) one can obtain \cite{Ake09} the kernel
\begin{equation}\label{Ksop}
 {\cal K}_N(z_1,z_2) = \sum_{k,l=1}^N P_{k-1}(z_1)\frac{Z^{-1}_{kl}}{r_{l-1}} P_{l-1}(z_2)
\end{equation}
expanded in terms of skew orthogonal monic polynomials $P_{k-1}(z)$ defined by
\begin{equation}\label{Psop}
  \int {\rm d}^2z_1\ {\rm d}^2z_2 {\cal F}(z_1, z_2) P_{k-1}(z_1)  P_{l-1}(z_2)= Z_{kl}\  r_{l-1}=r_{k-1}\ Z_{kl} \ .
\end{equation}
For $\tau>0$, (\ref{Ksop}) was also obtained directly from the jpdf in \cite{For08}. In this case the skew-orthogonal polynomials can be expressed \cite{For08} in terms of the Hermite polynomials $P_{2n}(z)=p_{2n}(z)$ and $P_{2n+1}=p_{2n+1}(z)-2n p_{2n-1}(z)$ with $p_n(z)= ({\tau}/{2})^{n/2} H_n({z}/{\sqrt{2\tau}})$ being the scaled Hermite polynomials as in Section \ref{sec4}, and with
$r_n=r_{n+1}=2\sqrt{2\pi}\, n!\, (1+\tau)$ for $n$ even.

In the qu-r case the saddle point integral representation of the kernel following from its relation to the average of characteristic polynomials is more involved, however the kernel can again be expanded in skew orthogonal polynomials (for $\tau >0$) [Kan02]. The resulting expression is the same as in (\ref{Ksop}) only now $P_{2n}(z)=\sum_{l=0}^{n}\frac{ 2^n n!}{2^l l!}p_{2l}(z)$ and $P_{2n+1}=p_{2n+1}(z)$, with the polynomials $p_n$ as before and
the constants $r_n$ in (\ref{Psop}) given by
$r_n=r_{n-1}=  2\pi\ n!\ (1-\tau)$ for $n$ odd.
Note that the expressions for $r_n$ give immediately the normalization constant in (\ref{jpdfR}): $C_{N,\tau}=(1+\tau)^{-N/2}C_N$ in the real case and and $C_{N,\tau}=(1-\tau)^{-N/2}C_N$ in the qu-r case with $C_N$ as in (\ref{cs}).

Having an explicit form for the kernel, all correlations, including in the real case the real-real and real-complex ones, can be found  \cite{For08,Bor08,Som08}. 

{\bf Strongly non-Hermitian limit}\quad
In this limit $u,v$ are assumed to be of order $\sqrt{N}$ as $N\to\infty$. We are going to evaluate the kernel ${\cal K}_N(u,v)$ (\ref{CP}) by the saddle point analysis of $P_{N-2}(u,v)$. Let us start with the circular real case (\ref{Puv4}) which is the simplest. Then the saddle point equations are
$a^* = \frac{Na^*}{uv+aa^*}$ and $a =\frac{Na}{uv+aa^*}$
with two solutions $a=a^* =0$ and $aa^*= N-uv$.
The first saddle-point  yields $P_{N-2}\simeq (uv)^{N-2}/(1-N/uv)$ and the second, which is actually a manifold, yields $P_{N-2}\simeq N^{N-2}\ \sqrt{2\pi N} \ {\rm e}^{uv-N}$. Thus the second is dominating for ${\rm Re} (uv) < N$ and otherwise the first is (excluding the neighborhood of $N=uv$). Hence, the $1$-point density $R_1(z)$ is asymptotically $R_1(z)= 1/\pi$ for $|z|<\sqrt{N}$ and is exponentially small outside this circle. Also, this gives immediately the asymptotic form of the kernel in the circular case in the bulk (${\rm Re} (z_1z_2) < N$):
\begin{equation}   \label{Kcr}
    {\cal K}_N(z_1,z_2) \simeq (2\sqrt{2\pi})^{-1}\, (z_1-z_2)\, {\rm e}^{z_1z_2}\quad \text{as $N\to\infty$.}
\end{equation}
Obviously, (\ref{Kcr}) could have been obtained directly from (\ref{10}). Here we have determined in addition the region of validity. The correlations ($n$-point densities) are then given with this formula in the region $|z_j|<\sqrt{N}$ , $j=1,2,\ldots, n$. Hence we call it the circular real case.

By similar reasoning we find in the circular qu-r case
\begin{equation}   \label{Kcq}
    {\cal K}_N(z_1,z_2) \simeq (2\pi (z_2-z_1))^{-1}\, {\rm e}^{z_1z_2}\quad \text{as $N\to\infty$,}
\end{equation}
for ${\rm Re} (z_1z_2) < N$  with both $z_1$ and $z_2$ (and also $z_1-z_2$) being of order $\sqrt{N}$. This implies the circular law in the qu-r case: $R_1(z)\simeq 1/\pi$ inside the circle $|z|=\sqrt{N}$ and away from the real line ($\im z \sim \sqrt{N}$) and $R_1(z)\simeq 0$ outside.

For the elliptic real case the relevant saddle point leads to the condition for the elliptic support  of $R_1(z)$. This has been found numerically and analytically using the replica trick already in the early paper \cite{Som88}.
Similarly, for the elliptic qu-r case we expect it can be shown by a saddle point analysis that in the large $N$-limit $R_1(z)$ is constant inside the ellipse with main half-axes $\sqrt{N}(1\pm\tau)$ along the $x$-and $y$- direction, and zero outside.

Having the expressions (\ref{Ksop}) for the kernel in terms of skew orthogonal polynomials, one can find the scaling limit of the kernel and the correlations for $N$ to infinity considering $z_{1,2}$ as being of order $1$ and letting the edge going to infinity. Since the 1-point density is constant it is not necessary to unfold. It turns out that in the real case  for $\tau=0$ this limit ${\cal K}_{\infty}(z_1,z_2)$ is already given by expression (\ref{Kcr}). For $\tau>0$ ${\cal K}_{\infty}(z_1,z_2)$ has been calculated in \cite{For08}
\begin{equation}   \label{Kcrtau}
    {\cal K}_{\infty}(z_1,z_2) =\frac{z_1-z_2}{2\sqrt{2\pi}(1-\tau^2)^{3/2}}\exp\left(\frac {z_1z_2}{1-\tau^2}-\frac{\tau(z_1^2+z_2^2)}{2(1-\tau^2)}\right)
\end{equation}
with the normalization corrected.
Obviously this is also valid for $-1<\tau<0$.
In the circular qu-r case ($\tau=0$) the asymptotic form of the kernel has been calculated in \cite{Kan02} and is already found in \cite{Meh04} in a somewhat different form.
Scaling it appropriately one obtains for the elliptic qu-r case
\begin{equation}   \label{Kcqtau}
    {\cal K}_{\infty}(z_1,z_2) =\frac{1}{2\sqrt{2\pi}(1-\tau^2) } \exp\left(\frac{z_1^2+z_2^2}{2(1+\tau)}\right) {\rm erf}\left(\frac{z_1-z_2}{\sqrt{2(1-\tau^2)}}\right) \  \ .
\end{equation}
Both expressions (\ref{Kcrtau}, \ref{Kcqtau}) yield the  bulk density $R_1(z)\simeq 1/\pi(1-\tau^2)$ corresponding to an elliptic support with main half axes $\sqrt{N}(1\pm \tau)$. In the real case the density of real eigenvalues is asymptotically again constant $R_1^R(x)\simeq 1/\sqrt{2\pi(1-\tau^2)}$ which together with the support $|x|<\sqrt{N}(1+\tau)$ gives a number of real eigenvalues of $\sqrt{2N(1+\tau)/\pi(1-\tau)}$ (cf (\ref{EKS})) \cite{For08}.

{\bf Weakly non-Hermitian limit}\quad
Finally we consider the limits  of weak non-Hermiticity \cite{Fyo97}. Here we put $\tau=1-a^2$ with $1-\tau$ being of order $1/N$ and let $N$ go to infinity assuming $z_i$ to be in the neighborhood of the origin. In the real case in this limit the kernel is given by \cite{For08}
\begin{equation}   \label{weakrK}
{\cal K}_{N}(z_1,z_2)\simeq \frac{N}{2\pi }\int_0^{1}{\rm d}u\ u\ \exp({-Na^2u^2})\sin(\sqrt{N}\ u(z_1-z_2))\, ,
     \end{equation}
and $f_{\tau}^2(z) \simeq \erfc(\frac{|z-z^*|}{2a})$. In the qu-r case $f_{\tau}^2(z) \simeq \rme^{\frac{(z-z^*)^2}{4 a^2}}$ and \cite{Kan02}
\begin{equation}   \label{weakqK}
{\cal K}_{N}(z_1,z_2)\simeq  \frac{\pi^{3/2}}{4a^3N^{3/2}}\int_0^{1}\frac{{\rm d}u}{ u}\ \exp({-Na^2u^2})\sin(\sqrt{N}\ u(z_1-z_2))\, .
     \end{equation}
As in the complex elliptic ensemble, it is convenient to scale $z_1, z_2, a $ and ${\cal K}_N$ with the local mean level spacing (here $=\pi/\sqrt{N}$ since the ensembles go in this limit to GOE/GSE with the semicircular density of eigenvalues $\rho_{sc}(x)= \frac{1}{\pi}\sqrt{N-{x^2}/{4}}$). Then one obtains a universal form of the correlations. For example, on the large $x$-scale ($z=x+\rmi y)$ the $1$-point density can be written as $R_1(z)\simeq \rho_{sc}(x)^2 P(\rho_{sc}(x)y,\rho_{sc}(x)a)$ with the probability density function
\[
P(y,a)=\delta(y)\int_0^{1}{\rm d}u\ \rme^{-\pi^2 a^2 u^2} + {\pi} \erfc\left(\frac{|y|}{a}\right)\int_0^{1}{\rm d}u\ u\ \rme^{-\pi^2 a^2 u^2}\sinh(2\pi u |y|)
\]
in the real case (GOE limit) \cite{Efe97} and in the qu-r case (GSE limit) \cite{Kol99}
\[
P(y,a)= \frac{y}{\pi^{3/2}a^3}{\exp}\left(-\frac{y^2}{a^2}\right)\int_0^{1}\frac{{\rm d}u}{u}\ \exp({-\pi^2 a^2 u^2})\sinh(2\pi u y)\, .
\]
Note that in the qu-r case despite the fact that the complex eigenvalue density vanishes exactly on the real axis in the limit pairs of complex conjugate eigenvalues approach the real axis and collapse giving Kramers degeneracy for the Hermitian ensemble with symplectic symmetry. In the real case  despite the fact that the density of real eigenvalues becomes constant, but very low, complex eigenvalues approach the real axis and in the limit give rise to the Wigner semicircle density which can be considered as a projection of the elliptic law in the  complex plane onto the real axis.

{\sc Acknowledgements}: H-JS acknowledges support by SFB/TR12 of the Deutsche Forschungsgemeinschaft. Gernot Akemann, Yan Fyodorov, Eugene Kanzieper and Dmitry Savin are thanked for helpful comments on earlier versions of this manuscript.


\begin{thebibliography}{Abc84a}


\bibitem[Ake03]{Ake03} G. Akemann, G. Vernizzi,
Nucl. Phys. B {\bf 660 [FS]} (2003) 532

\bibitem[Ake05a]{Ake05} G. Akemann,
Nucl. Phys. B {\bf 730} (2005) 253

\bibitem[Ake05b]{Ake05a} G. Akemann, J.C. Osborn, K. Splittorf, J.J.M. Verbaarschot,
Nucl. Phys. B {\bf 712} (2005) 287

\bibitem[Ake07]{Ake07} G. Akemann, F. Basile,
Nucl. Phys. B {\bf 766} (2007) 766

\bibitem[Ake09a]{Ake09} G. Akemann, M. J. Phillips, H.-J. Sommers,
J. Phys. A 
{\bf 42} (2009) 012001

\bibitem[Ake09b]{Ake09a} G. Akemann, M.J. Phillips, L.Shifrin,
J. Math. Phys. {\bf 50} (2009) 063504

\bibitem[Ake09c]{Ake09b} G. Akemann, M.J. Phillips, H.-J. Sommers,
arXiv:cond-mat/0911.1276v1


\bibitem[Ame08]{Ame08} Y. Ameur, H. Hedenmalm, N. Makarov, arXiv:08070375 [math.PR]

\bibitem[Ben09]{Ben08} M. Bender,
Probability Theory and Related Fields (2009) doi: 10.1007/s00440-009-0207-9 (arXiv:0808.2608 [math.PR])


\bibitem[Ber02]{Ber02} D. Bernard, A. LeClair,
J. Phys A {\bf 35} (2002) 2555; also arXiv:cond-mat/0110649

\bibitem[Bor09]{Bor08} A. Borodin, C.D. Sinclair,
Commun. Math. Phys.{\bf 291} (2009) 177

\bibitem[Bru09]{Bru09} W.Bruzda, V.Cappelini, H.-J.Sommers, K.\.Zyczkowski,
Phys. Lett. A {\bf 373} (2009) 320

\bibitem[DiF94]{DiF94} F. Di Francesco, M. Gaudin, C. Itzykson, F. Lesage
Int. J. Mod. Phys. A {\bf 9} (1994) 4257

\bibitem[Ede97]{Ede97} A. Edelman,
J. Multivar. Anal. {\bf 60} (1997), 203

\bibitem[Ede94]{Ede94} A. Edelman, E. Kostlan, M. Shub,
J. Amer. Math. Soc. {\bf 7} (1994), 247

\bibitem[Efe97]{Efe97} K.B. Efetov,
Phys. Rev. Lett.\  {\bf 79} (1997), 491

\bibitem[Fei97]{Fei97} J. Feinberg, A. Zee, Nucl. Phys. B {\bf 501} (1997) 643; 
J. Feinberg, J. Phys. A {\bf 39} (2006) 10029.


\bibitem[For97]{For97} P.J. Forrester, B. Jancovici,
Int. J. Mod. Phys. A {\bf 11} (1997) 941

\bibitem[For99]{For99} P.J. Forrester, G. Honner,
J. Phys. A 
{\bf 32}, (1999), 2961

\bibitem[For07]{For07} P.J. Forrester, T. Nagao,
Phys. Rev. Lett. {\bf 99}, (2007), 050603

\bibitem[For08]{For08} P.J. Forrester, T. Nagao,
J. Phys. A
{\bf 41} (2008), 375003

\bibitem[For09a]{For09a} P.J. Forrester, A. Mays,
J. Stat. Phys. {\bf 134} (2009), 443

\bibitem[For09b]{For09b} P.J. Forrester, E. M. Rains,
J. Phys. A {\bf 42} (2009) 385205

\bibitem[Fyo97]{Fyo97} Y.V. Fyodorov, B.A. Khoruzhenko, H.-J. Sommers,
Phys.\ Lett.\ {\bf A226} (1997) 46; Phys. Rev. Lett. {\bf 79} (1997) 557

\bibitem[Fyo98]{Fyo98} Y.V. Fyodorov, B.A. Khoruzhenko, H.-J. Sommers,
Ann. Inst. H. Poincar\'e Phys. The\'or.  {\bf 68}  (1998) 449

\bibitem[Fyo99]{Fyo99} Y.V. Fyodorov, B.A. Khoruzhenko,
Phys. Rev. Lett. {\bf 83} (1999) 65

\bibitem[Fyo03]{Fyo03} Y.V. Fyodorov, H.-J. Sommers,
J. Phys. A
{\bf 36} (2003) 3303

\bibitem[Fyo07]{Fyo07} Y.V. Fyodorov, B.A. Khoruzhenko, Commun. Math. Phys. {\bf 273} (2007) 561

\bibitem[Gar02]{Gar02} A.M. Gar\'cia-Gar\'cia, S.M. Nishigaki, J.J. Verbaarschot,
Phys. Rev. E {\bf 66} (2002) 016132

\bibitem[Gin65]{Gin65} J. Ginibre,
J. Math. Phys. {\bf 6} (1965) 440

\bibitem[Gir85]{Gir85} Girko V.L.,
Theor. Prob.  Appl. {\bf 29} (1985) 694

\bibitem[Gir86]{Gir86} Girko V.L.,
Theor. Prob.  Appl.
{\bf 30} (1986) 677

\bibitem[Gro88]{Gro88} R. Grobe, F. Haake, H.-J. Sommers,
Phys. Rev. Lett. {\bf 61} (1988) 1899

\bibitem[Jia06]{Jia06} T. Jiang,
Ann. Prob. {\bf 34} (2006) 1497

\bibitem[Kwa06]{Kwa06} J.Kwapien, S. Drozdz, A.Z. Gorski, F. Oswiecimka, Acta Phys. Pol. B {\bf 37} (2006) 3039

\bibitem[Leh91]{Leh91} N. Lehmann, H.-J. Sommers,
Phys. Rev. Lett. {\bf 67} (1991) 941

\bibitem[May72]{May72} R.M. May,
Nature {\bf 298} (1972) 413

\bibitem[Meh04]{Meh04} M.L. Mehta, {\it Random Matrices}, 3rd ed., Academic Press, 2004.

\bibitem[Kan02]{Kan02} E. Kanzieper,
J. Phys. A
{\bf 35} (2002) 6631

\bibitem[Kan05a]{Kan05a} E. Kanzieper,
In: Frontiers in Field Theory, ed. O. Kovras, 2005 Nova Science Publ.\ pp. 23--51.

\bibitem[Kan05b]{Kan05b} E. Kanzieper, G. Akemann
Phys. Rev. Lett. {\bf 95} (2005) 230201;
G. Akemann, E. Kanzieper,
J. Stat. Phys. {\bf 129} (2007), 1159

\bibitem[Kol99]{Kol99}	A.V. Kolesnikov, K.B. Efetov
Waves Ran. Media  {\bf 9} (1999) 71

\bibitem[Mag08]{Mag08}  U. Magnea,
J. Phys. A {\bf 41} (2008) 045203

\bibitem[Oas97]{Oas97} G. Oas
Phys. Rev. E {\bf 55} (1997) 205

\bibitem[Osb04]{Osb04} J. C. Osborn,
Phys. Rev. Lett. {\bf 93} (2004) 222001

\bibitem[Som88]{Som88} H.-J. Sommers, A. Crisanti, H. Sompolinsky, Y. Stein,
Phys. Rev. Lett.  {\bf 60} (1988) 1895

\bibitem[Som07]{Som07} H.-J. Sommers
J. Phys. A 
{\bf 40} (2007) F671

\bibitem[Som08]{Som08} H.-J. Sommers, W. Wieczorek,
J. Phys. A
{\bf 41} (2008) 405003

\bibitem[Som09]{Som09} H.-J. Sommers, B.A. Khoruzhenko,
J. Phys. A {\bf 42} (2009) 222002 

\bibitem[Tao08]{Tao08} T. Tao, V. Vu, M. Krishnapur,
arXiv:0807.4898 [math.PR]

\bibitem[Tra98]{Tra98} C. Tracy, H. Widom,
J. Stat. Phys. {\bf 92} (1998) 809

\bibitem[Zyc00]{Zyc00} K. \. Zyczkowski, H.-J. Sommers,
J. Phys. A
{\bf 33} (2000) 2045

\end{thebibliography}
\end{document}